# Unscrambling Structured Chirality with Structured Light at Nanoscale Using Photo-induced Force

Mohammad Kamandi, Mohammad Albooyeh, Mehdi Veysi, Mohsen Rajaei, Jinwei Zeng, Kumar Wickramasinghe and Filippo Capolino*

Department of Electrical Engineering and Computer Science, University of California, Irvine, California 92697, USA

## ABSTRACT

We introduce a microscopy technique that facilitates the prediction of spatial features of chirality of nanoscale samples by exploiting the photo-induced optical force exerted on an achiral tip in the vicinity of the test specimen. The tip-sample interactive system is illuminated by structured light to probe both the transverse and longitudinal (with respect to the beam propagation direction) components of the sample's magnetoelectric polarizability as the manifestation of its sense of handedness, *i.e.*, chirality. We specifically prove that although circularly polarized waves are adequate to detect the transverse polarizability components of the sample, they are unable to probe the longitudinal component. To overcome this inadequacy and probe the longitudinal chirality, we propose a judiciously engineered combination of radially and azimuthally polarized beams, as optical vortices possessing pure longitudinal electric and magnetic field components along their vortex axis, respectively. The proposed technique may benefit branches of science like stereochemistry, biomedicine, physical and material science, and pharmaceutics.

## 1) INTRODUCTION

Structure determination of chiral specimens is of great interest since the fundamental building blocks of life, *i.e.*, proteins and nucleic acids are built of chiral amino acids and chiral sugar [1]. Indeed, the optical activity of chiral structures, is a key parameter in molecular identification techniques to recognize the type of a molecule or to determine its structure thanks to the discriminatory behavior of chiral molecules in interaction with the incident light possessing distinct sense of polarization [2,3]. For instance, for a protein, determining the structure refers to resolving its four levels of complexity, *i.e.*, primary, secondary, tertiary and quaternary which defines not only the sequence of amino acids but also reveals the three dimensional arrangement of atoms in that protein [4]. This information is of supreme importance in modifying and utilizing proteins for new purposes such as creating protein-based anti-body drug conjugates for cancer treatment or modifying the proteins in the bread [5–9]. To determine the structure of chiral samples such as protein, noninvasive spectroscopic techniques based on optical rotation (OR), circular dichroism (CD), and Raman optical activity (ROA) have been proposed and vastly studied [10–13]. In these methods, owing to the optical activity of chiral structures, the difference between the absorbed left-hand and right-hand circularly polarized (CP) light is measured and not only the chirality but also some important information about the structure of chiral sample is obtained [14–16]. Specifically using CD, not only the primary structure of a protein, but also its secondary structure is determined. Moreover, the CD spectrum of a protein might give the fingerprint of the tertiary structure. However, the main limitations of this method are as following: (i) it is unable of providing high resolution structural details and (ii) it demands considerable amount of the material for detection. These limitations mainly originate from the fact that CD measures the averaged far-field radiation which misses the essential information only carried in near field, thus, this dearth calls for possible potent near-field measurement techniques which are more promising for providing nanoscale details.

The capability of atomic force microscopy (AFM) which was originally introduced to derive the morphological properties of the specimen [17,18] has been recently expanded to measure optical properties of the specimen by exciting the tip-sample interactive system with an incident light beam in the so called photo-induced force microscopy (PiFM) [19–22]. In PiFM the tip-sample system is illuminated by an electromagnetic field and the force exerted on the tip in the vicinity of the sample is measured and used to perform linear and non-linear spectroscopy to obtain optical characteristics of the sample at the nanoscale [23]. Unlike conventional spectroscopy techniques the PiFM utilizes the near-field data from the interaction between the tip-sample interactive system and light, and hence, is not limited by diffraction[24,25] and has high signal to noise ratio (SNR)[25,26]. Recently, PiFM has been utilized to report the chirality and enantiomer-type[1] of a chiral sample theoretically, [27] with an achiral tip, and experimentally, [28] with a spiral tip.

In this paper, by taking the advantage of PiFM, we introduce a new technique to determine the longitudinal and transverse chirality components of specimens with high resolution. Unlike CD, this method has high spatial resolution of sub 100-nm-scale. In particular, we calculate the exerted force on the tip in the vicinity of a test sample to identify the different components of the sample handedness which are identified through the magnetoelectric polarizability (or equivalently chirality parameter). Specifically, we prove that proper excitations for detection of *transverse* (with respect to the propagation direction) components of sample chirality in our proposed method are CP waves since they possess field components transverse to the direction of propagation. Despite this capability, we unravel the failure of CP waves in detecting the longitudinal (*i.e.*, along the propagation direction) component of sample chirality. We particularly prove that a light beam with longitudinal electric or (and) magnetic field component(s) would be an appropriate candidate for the detection of longitudinal chirality component. In this context, optical vortices with longitudinal electric or magnetic field component along the vortex axis[29] serve as suitable practical excitations. With the goal of detection of longitudinal component of chirality and in order to optimize the interaction between the chiral sample and the excitation, we propose a combination of an azimuthally and a radially polarized beam (APB and RPB)[30–32] with an appropriate phase difference as the illumination. As we prove, our proposed technique for the excitation and probing longitudinal chirality component provides a fundamental advantage in experiments when we are limited to illumination from one side which is common and applied in most PiFM cases.[32]

The paper is structured as follows: In Sec. II we demonstrate the operation principle of our proposed method by applying the dipolar approximation technique. Moreover, we discuss the magnetoelectric polarizability and its relation with the terminology of chiral bianisotropic material. In Sec. III we exhibit CP excitations to detect the transverse components of the sample handedness (or equivalently magnetoelectric polarizability), and show how they fail in probing the longitudinal chirality component. In Sec. IV we bring up the idea of superposition of an APB and an RPB as a suitable excitation scenario for detecting the longitudinal component of the sample handedness which was unrecognizable under the CP excitation. Lastly, we conclude the paper with some remarks.

---

[1] Enantiomer refers to each pairs of chiral samples that are mirror images of each other.

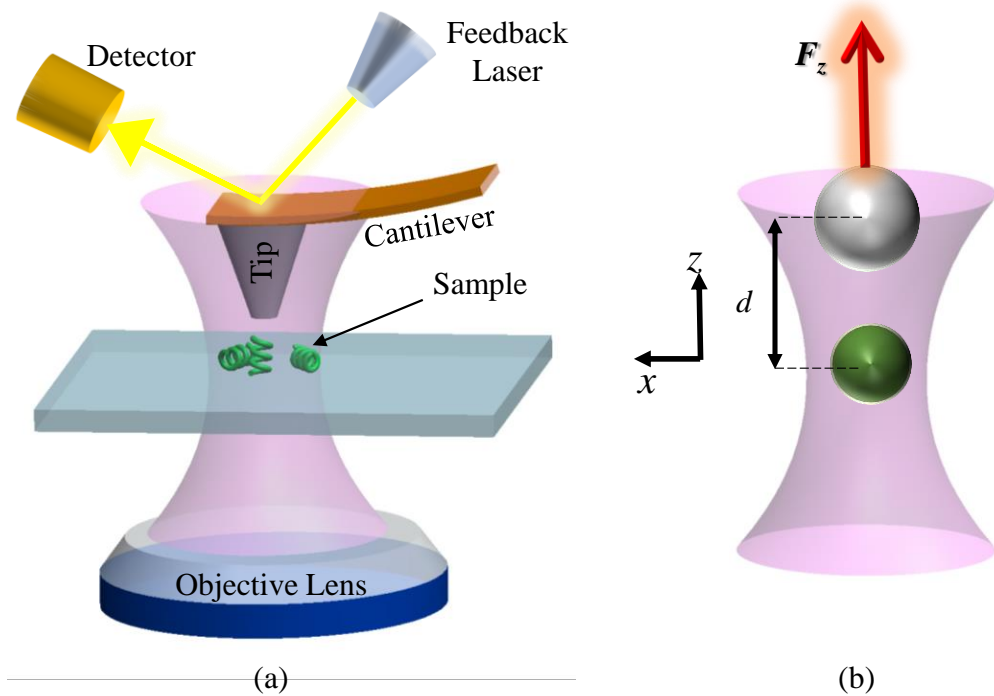

FIG. 1. (a) Schematic of photo-induced force microscopy capable of probing different components of the handedness of a chiral sample particle. (b) Simplified model where the tip-apex and sample are represented by nanospheres with electric (and perhaps magnetic) dipole moment(s). We investigate electromagnetic forces due to near-field interaction between the two nanostructures, the plasmonic tip and the chiral sample, under external illumination with structured chiral light. In this paper, the transverse components refer to the $x$ and $y$ directions whereas the longitudinal one refers to the $z$ direction.

## 2) OBJECTIVE AND OPERATIONAL PRINCIPLE

Figure 1 depicts the PiFM set-up which is utilized throughout this paper for both the transverse and longitudinal chirality investigations. As it is shown, the interactive system composed of a nanoscale chiral sample and the microscope tip is illuminated by incident focused light propagating along the positive $z$ direction (referred here as the longitudinal direction) from the bottom side through an objective lens. The incident beam induces electric and perhaps magnetic polarization currents on both the sample and tip-apex which consequently provide secondary scattered electromagnetic waves. In the PiFM technique, the sample and tip-apex are supposed to be close enough (compared to the operational wavelength), so that they interact through each other's near-zone scattered fields. That is why a noticeable force is exerted on the tip-apex due to the transfer of momentum from the scattered photons. In this paper, we are specifically interested in the force exerted on the tip-apex in the longitudinal direction ($z$-direction in Fig. 1) since it can be measured using a PiFM [32]. Moreover, as discussed previously, unlike conventional chiroptical techniques such as CD for detecting chirality, PiFM is capable of detecting chirality at nanoscale since it exploits the information carried in the near-field scattered by the sample.

We assume both tip-apex and sample to be small compared to the wavelength of incident light so that their main interaction feature is modeled via dipoles' scattering. In our theoretical analysis and for the proof of concept, we approximate the tip-apex and sample as two nanospheres and model the tip-apex by equivalent electric and magnetic dipoles, as already done in Refs [23,26,27,33], and the chiral sample by its bianisotropic polarizability and use the dipole approximation to investigate their interaction (see Fig 1 (b)). Note that this dipolar approximation has successfully been exploited to model photo-induced force microscopy in Refs. [23,26,27,33] to extract optical properties of samples. Here, we model the sample with its photo-induced electric and magnetic dipole moments $p$ and $m$, respectively, given by [34–36]

$$\begin{bmatrix} p_s \\ m_s \end{bmatrix} = \begin{bmatrix} \underline{\alpha}_s^{ee} & \underline{\alpha}_s^{em} \\ \underline{\alpha}_s^{me} & \underline{\alpha}_s^{mm} \end{bmatrix} \cdot \begin{bmatrix} E^{loc}(r_s) \\ H^{loc}(r_s) \end{bmatrix}, \qquad (1)$$

where subscript "s" represents the "sample" and $E^{loc}$ and $H^{loc}$ are, respectively, the local electric and magnetic field vectors acting on the sample at position $r_s$. Moreover, $\underline{\alpha}^{ee}$, $\underline{\alpha}^{mm}$, $\underline{\alpha}^{em}$, and $\underline{\alpha}^{me}$ are second rank tensors describing the electric, magnetic, magnetoelectric and electromagnetic *polarizabilities* of the sample, respectively. The last two tensors are describing bianisotropy of the sample particle, *i.e.*, they represent the electric (magnetic) response of the particle to the magnetic (electric) excitation [34,35,37]. Note that here the magnetic dipole moment is defined as $m = \mu_0 \int_V dv\, r \times J / 2$, with $J$ and $r$ being the volumetric current density and the position vector in volume $V$, respectively, and $\mu_0$ is the vacuum permeability. Using this definition (that includes $\mu_0$) provides the same units for the electromagnetic and magnetoelectric polarizabilities $\underline{\alpha}^{em}$, and $\underline{\alpha}^{me}$ and is chosen following Ref.[27]. Before delving into the physical details of our method for probing transverse and longitudinal handedness of chiral particles, it is useful to differentiate chiral nanoparticles from other types of bianisotropic nanoparticles by discussing the properties of magnetoelectric and electromagnetic polarizability tensors. All bianisotropic particles may be classified into four types: chiral, omega, Tellegen, and "moving". While the first two fall into the category of reciprocal particles, the last two types exhibit nonreciprocity properties. For reciprocal particles the magnetoelectric and electromagnetic polarizability tensors reduce to one since the property $\underline{\alpha}^{me} = -\left(\underline{\alpha}^{em}\right)^T$ holds, whereas the electric and magnetic polarizability tensors satisfy the properties $\underline{\alpha}^{ee} = \left(\underline{\alpha}^{ee}\right)^T$ and $\underline{\alpha}^{mm} = \left(\underline{\alpha}^{mm}\right)^T$, and $T$ represents the tensor transpose[34]. To summarize, a chiral particle must possess two specific properties in terms of the induced moment and the incident field: a) it should be bi-isotropic (or bianisotropic), that is, the *electric* (*magnetic*) dipole response must be induced by the incident *magnetic* (*electric*) field, and b) the induced electric and magnetic dipole moments $p$ and $m$ must be *parallel*. The magnetoelectric polarizability tensor of a particle in Cartesian coordinates is represented by the matrix

$$\underline{\underline{\alpha}}^{em} = \begin{bmatrix} \alpha_{xx}^{em} & \alpha_{xy}^{em} & \alpha_{xz}^{em} \\ \alpha_{xx}^{em} & \alpha_{yy}^{em} & \alpha_{yz}^{em} \\ \alpha_{zx}^{em} & \alpha_{zy}^{em} & \alpha_{zz}^{em} \end{bmatrix}. \tag{2}$$

In this paper, we investigate reciprocal chiral particles, that is, magnetoelectric polarizability is diagonal, having only $\alpha_{xx}^{em}$, $\alpha_{yy}^{em}$ or $\alpha_{zz}^{em}$ as non-vanishing components. Since the propagation direction of the incident field in our set-up is along the $z$-axis, in the following we refer to the first two polarizability components $\alpha_{xx}^{em}$ and $\alpha_{yy}^{em}$ as the *transverse* components and the last component $\alpha_{zz}^{em}$ as the *longitudinal* one.

As we discussed, the goal is to characterize the transverse and longitudinal magnetoelectric polarizabilities of a chiral nanoparticle by exploiting different structured light illumination scenarios and by exploring the exerted force on the tip in the vicinity of the chiral sample. To that end, we consider an achiral tip-apex and use the dipolar approximation to model its response to an arbitrary electromagnetic wave as

$$\mathbf{p}_t = \underline{\underline{\alpha}}_t^{ee} \cdot \mathbf{E}^{loc}(\mathbf{r}_t), \qquad \mathbf{m}_t = \underline{\underline{\alpha}}_t^{mm} \cdot \mathbf{H}^{loc}(\mathbf{r}_t). \tag{3}$$

Here, the local fields $\mathbf{E}^{loc}$ and $\mathbf{H}^{loc}$ (which include the contributions from both the *incident* field and the *scattered* field from the sample) are calculated at the tip-apex position $\mathbf{r}_t$, where subscript "t" denotes the "tip". Note that in our set-up, we employ an achiral tip-apex, *i.e.*, $\underline{\underline{\alpha}}_t^{me} = \underline{\underline{\alpha}}_t^{em} = 0$. Accordingly, the local electric and magnetic fields at the tip position are described by

$$\begin{aligned} \mathbf{E}^{loc}(\mathbf{r}_t) &= \mathbf{E}^{inc}(\mathbf{r}_t) + \underline{\underline{G}}^{EE}(\mathbf{r}_t, \mathbf{r}_s) \cdot \mathbf{p}_s + \underline{\underline{G}}^{EM}(\mathbf{r}_t, \mathbf{r}_s) \cdot \mathbf{m}_s, \\ \mathbf{H}^{loc}(\mathbf{r}_t) &= \mathbf{H}^{inc}(\mathbf{r}_t) + \underline{\underline{G}}^{ME}(\mathbf{r}_t, \mathbf{r}_s) \cdot \mathbf{p}_s + \underline{\underline{G}}^{MM}(\mathbf{r}_t, \mathbf{r}_s) \cdot \mathbf{m}_s. \end{aligned} \tag{4}$$

Here $\underline{\underline{G}}^{EE}$ and $\underline{\underline{G}}^{ME}$ are the dyadic Green's functions that provide the electric and magnetic fields due to an electric dipole $\mathbf{p}_s$, whereas $\underline{\underline{G}}^{EM}$ and $\underline{\underline{G}}^{MM}$ are the dyadic Green's functions that provide the electric and magnetic fields due to a magnetic dipole $\mathbf{m}_s$, respectively [38,39].

The general expression of the time-averaged optical force exerted on the tip reads [40–44]

$$F = \frac{1}{2} \operatorname{Re} \left[ \left( \nabla E^{loc}(\mathbf{r}_t) \right)^* \cdot p_t + \left( \nabla H^{loc}(\mathbf{r}_t) \right)^* \cdot m_t - \frac{ck^4}{6\pi} \left( p_t \times m_t^* \right) \right], \tag{5}$$

in which $c$ is the speed of light and $k$ is the ambient wavenumber, the asterisk represents complex conjugation and $\nabla$ is the gradient operator which its $ij$-th component, when applied to a vector A is equal to $\partial_i A_j$ (here, $i, j = x, y, z$ in Cartesian coordinates and $\partial_i$ is partial derivative with respect to the $i$-th spatial coordinate. See more details in the supplemental materials, and in Refs.[26,27]). Moreover, we assume that every field is monochromatic and the time dependence $\exp(-i\omega t)$ is assumed and suppressed, and the International System of Units (SI) is utilized throughout the paper.

In this paper we employ right and left handed CP Gaussian beams as excitation for probing the transverse component of the sample's magnetoelectric polarizability. However, we prove that the CP illumination scenario is unable of identifying the longitudinal component of the magnetoelectric polarizability. Consequently, for the detection of longitudinal component of the magnetoelectric polarizability, we propose an alternative illumination scheme based on a combination of an APB and RPB with a proper phase difference. The reason is that such combination *exclusively* interacts with the electric and magnetic dipoles of the sample in the *longitudinal* direction, interacting with the sample's longitudinal chirality, whereas CP waves that propagate along the longitudinal direction lack such characteristic.

Even a CP wave which is obliquely (with respect to the longitudinal direction) illuminating the tip-sample system would not be able to provide the same information. Although such obliquely incident CP wave is capable of interacting with the longitudinal dipoles of the sample, the detection of longitudinal chirality is not exclusively possible since both the longitudinal and transverse dipole components of the sample are excited and it is impossible to distinguish different chirality components with a couple of excitations from one side.

### 3)  PROBING THE TRANSVERSE HANDEDNESS OF CHIRAL SAMPLES

As discussed, CP beams are used as excitation in our proposed PiFM set-up. Specifically, we first send a right-hand CP (RCP) beam and then a left hand CP (LCP) beam and calculate the $z$-directed exerted force on the tip-apex in the vicinity of the sample for each case. For an achiral sample, the measured forces exerted on the tip-apex are equal for both excitation scenarios. However, owing to the optical activity of chiral materials and their rotational asymmetry, a chiral sample reacts differently under illuminations with opposite sense of handedness (here RCP and LCP). This distinction is manifested in the exerted force on the tip-apex. To verify it, we consider an example when the test sample and the plasmonic tip-apex are illuminated with a CP Gaussian beam propagating along the $z$-direction with a 1mW incident power at a wavelength of $\lambda = 400\ nm$. We assume that the waist of the Gaussian beam is $w_0 = 0.7\lambda$ (note that the actual waist of the beam is $2w_0$ [30]). Moreover, the sample is placed at the minimum waist $z$-plane of the beam (which hereafter we call beam waist) due to the higher strength of the field. For simplicity and in order to demonstrate the capability of our proposed method, we first assume that the sample is isotropic, *i.e.*, $\underline{\alpha}^{em} = \alpha^{em}\underline{I}$ and spherical. The chirality of the sample is denoted by the chirality parameter $\kappa$ which describes the degree of handedness of a material [27] and here we vary it from -1 to 1. The tip-apex is also assumed to be isotropic and spherical and both the sample and the tip-apex are considered to have equal radii $a_s = a_t = 50\ nm$ with center-to-center distance of $d = 110\ nm$ (see Fig. 2). Furthermore, without loss of generality and for the proof of concept, the relative permittivity of the sample is assumed to be $\varepsilon_s = 2.4$ whereas that of the plasmonic tip-apex is assumed to be $\varepsilon_t = -3.6 + i0.19$ (this is the relative permittivity of silver at the operational wavelength). In Fig. 2, we have depicted the $z$-component of the exerted force on the plasmonic tip-apex versus chirality parameter of the sample for both RCP and LCP incident waves using Eq. (5). The results clearly exhibit the potential of our proposed method in the detection of sample chirality at nanoscale. We define the time-averaged differential force along the $z$ direction as $\Delta F_z = F_z^{RCP} - F_z^{LCP}$, *i.e.*, the difference between the $z$-component of the exerted force on the tip-apex when the illumination is, respectively, an RCP and an LCP wave. This differential force is able to separate the effect of chirality of the sample from its other properties, *i.e.*, electric and perhaps magnetic (represented by the electric and magnetic polarizabilities). As it is illustrated in Fig. 2, the exerted differential force on the tip-apex for an achiral sample, *i.e.* $\kappa = 0$, is zero, however, as the magnitude of the chirality parameter of the sample grows, the differential force exerted on the tip-apex increases. Though in all the figures, results are obtained considering also the small force contribution due to the negligible magnetic dipole of the tip-apex as in Eq. (5) in the following we provide simple formulas obtained by neglecting such contribution. In this regard, the *z*-component of the general force expression of Eq. (5) reduces to[45,46]

$$F_z = \mathrm{Re}\left[ \left(\partial E_x^{\mathrm{loc}}(z)/\partial z\right)_{z=z_t}^* p_{\mathrm{t},x} + \left(\partial E_y^{\mathrm{loc}}(z)/\partial z\right)_{z=z_t}^* p_{\mathrm{t},y} \right]/2.$$ Furthermore, neglecting the field's phase

difference between the tip-apex and sample, due to their subwavelength distance, it is shown that the difference between the forces exerted on the tip-apex for two CP plane wave illuminations with opposite handedness reads[27,47,48]

$$\Delta F_z \approx -\frac{3|\mathrm{E}_0|^2}{4\pi\sqrt{\varepsilon_0\mu_0}\,d^4}\,\mathrm{Im}\left\{ \alpha_{\mathrm{t}}^{ee}\left(\alpha_{\mathrm{s}}^{em}\right)^* \right\}. \tag{6}$$

in which $\varepsilon_0$ is the vacuum permittivity and $d$ is the center-to-center distance between the sample and tip-apex. Note that in deriving this formula, we assume CP plane waves with electric field magnitude $|\mathrm{E}_0|$ on the beam axis since it simplifies our analytical calculations. Gaussian beams with a large beam waist can be reliably approximated by plane waves around their beam axis. Firstly, Eq. (6) clearly demonstrates that the differential force depends linearly on the electric polarizability of the tip-apex, hence, the material, size and shape of the tip-apex play crucial roles in determining the differential force. Specifically, in a previous study[27] the importance of the shape and material of the tip-apex has been further discussed[49,50]. For example, as discussed previously[27], by choosing a plasmonic tip-apex, we get stronger electric dipolar response and hence observe more pronounced differential force on the tip-apex. More importantly, Eq. (6) illustrates that the differential force is also linearly related to the magnetoelectric polarizability of the sample which in the quasi-static regime is approximated by[51,52]

$$\alpha_{\mathrm{s}}^{em} = 12i\pi a_{\mathrm{s}}^3 \sqrt{\varepsilon_0\mu_0}\, \frac{\kappa}{(\varepsilon_{\mathrm{s}}+2)(\mu_{\mathrm{s}}+2)-\kappa^2}. \tag{7}$$

where the subscript "s" represents "sample". After inserting Eq. (7) into Eq. (6) we observe that $\Delta F_z$ is linearly proportional to $\kappa$ since $(\varepsilon_{\mathrm{s}}+2)(\mu_{\mathrm{s}}+2)$ at the denominator is usually much larger than $\kappa^2$. Moreover, we observe that the sign of the differential force $\Delta F_z$ depends on the sign of the chirality parameter $\kappa$, which proves that our proposed method is capable of identifying the enantiomer type. Furthermore, it is also worth noting that $\Delta F_z$ is inversely related to the quartic power of the probe-sample distance.

It has been shown that using this method, with a beam power of 1mW and minimum waist radius related to $w_0 = 353\,\mathrm{nm}$, and a tip-apex with radius $a_{\mathrm{t}} = 60\,\mathrm{nm}$ and assuming a force of 0.1 pN as the instrument sensitivity[17,27], a sample with radius $a_{\mathrm{s}} = 70\,\mathrm{nm}$ and chirality of $\kappa = 0.04$ can be detected[27]. It is worth mentioning that a chirality parameter of the order of $\kappa \sim 10^{-2}$ which corresponds to a specific rotation[2] of $[\alpha]_D \sim 1000000°$ is still a large value compared to chirality parameters of abundant chiral molecules in nature such as Glucose ($C_6H_{12}O_6$) or Testosterone ($C_{19}H_{28}O_2$) which is in the order of $\kappa \sim 10^{-6}$ [53–56] (corresponding to specific rotation in the order of $[\alpha]_D \sim (100\text{-}200)°$). Indeed, only a few molecules and compounds such as Helicene or Norbornenone possess specific rotation angles of the order of $[\alpha]_D \sim 1000000°$ [57,58]. However, we emphasize that our technique, in contrast to conventional chiroptical techniques, is capable of detecting nanosamples in the order of ~50-100nm. The detection of chiral samples with smaller chirality parameter requires some extra considerations, like increasing the incident power which is possible in cryogenic conditions [59] or in liquids [60–62] or using nanotips with stronger electric polarizability, or even explore nanotips able to express magnetic response[49].

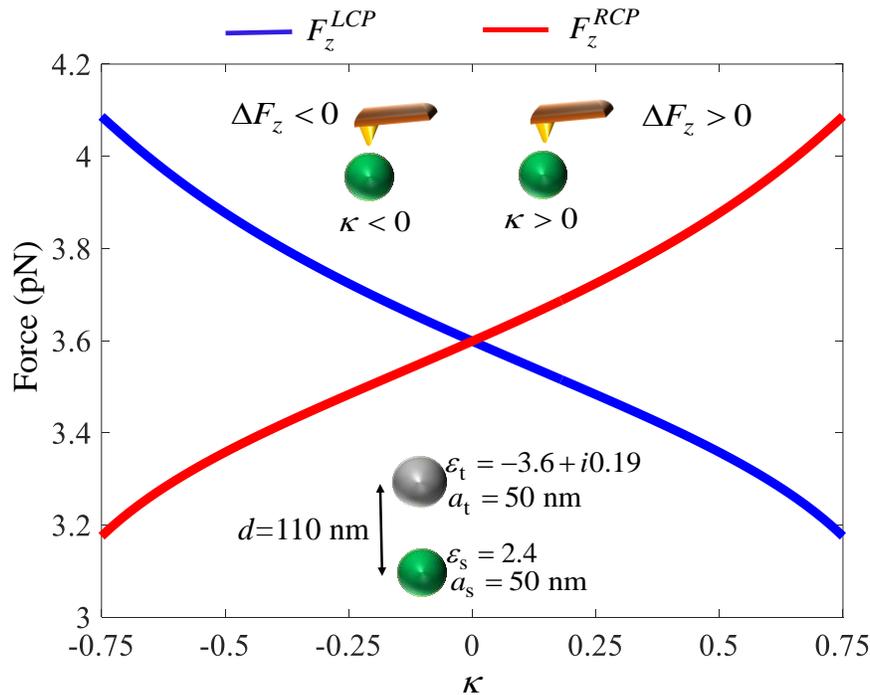

FIG. 2. The exerted force on the tip-apex for two incident scenarios of RCP and LCP illumination. Here, $a$ and $\varepsilon$ represent radius and permittivity, respectively.

So far, we have shown the ability of our proposed technique to detect and characterize isotropic chiral particles, *i.e*, particles with $\alpha_{xx}^{em} = \alpha_{yy}^{em} = \alpha_{zz}^{em}$ at nanoscale. However, as we have already discussed, it would be interesting to investigate the capability of our proposed excitation scenario (RCP/LCP waves) in distinguishing between the transverse and longitudinal components of the magnetoelectric polarizability. To that end, we note that the magnetoelectric polarizability tensor of the sample, in Cartesian coordinates is represented by the matrix

---

[2] Specific rotation is defined as the optical rotation in degrees per decimeter divided by the density of optically active material in grams per cubic centimeter

$$\underline{\alpha}_s^{em} = \begin{bmatrix} \alpha_{xx}^{em} & 0 & 0 \\ 0 & \alpha_{yy}^{em} & 0 \\ 0 & 0 & \alpha_{zz}^{em} \end{bmatrix}. \tag{8}$$

We assume here the exciting beam and the tip-apex have the same parameters as in the previous example related to Fig. 2. We study two different cases with samples' polarizabilities given by: 1) $\alpha_{xx}^{em} = \alpha_{yy}^{em} \neq 0$ and $\alpha_{zz}^{em} = 0$; and 2) $\alpha_{zz}^{em} \neq 0$ and $\alpha_{xx}^{em} = \alpha_{yy}^{em} = 0$. In this work we do not distinguish between the two transverse polarizability components $\alpha_{xx}^{em}$ and $\alpha_{yy}^{em}$, but we only distinguish between the transverse (*i.e.*, $\alpha_{xx}^{em}$ and $\alpha_{yy}^{em}$) and longitudinal ($\alpha_{zz}^{em}$) polarizability components of the sample nanoscatterer. In Fig.3 we have depicted the *z*-component of the differential force versus nonzero component of magnetoelectric polarizability for each case normalized to the magnetoelectric polarizability of the isotropic sample sphere studied in Fig. 2 with $\kappa = 0.75$, *i.e.* $\alpha_{iso}^{em}(\kappa = 0.75)$. (An analytic expression of such a polarizability for a chiral sphere is shown in Ref.[27] based on Ref.[63]) As it is clear, for a chiral sample with any nonzero *transverse* polarizability we can detect chirality with RCP and LCP illuminations since $\Delta F_z \neq 0$. However, for samples with only *longitudinal* magnetoelectric polarizability (with vanishing transverse magnetoelectric polarizability components) we are not able to characterize chirality with CP since $\Delta F_z = 0$. To clarify the reason of the failure of CP excitation in detecting longitudinal chirality component, we note that under paraxial approximation the *z*-components of the local fields, both at the tip-apex and sample location vanish for any excitation field that lacks the *z*-component (see supplemental materials). This means that the longitudinal component of the magnetoelectric polarizability does not contribute to the calculation of the electric and magnetic dipole moments in Eq. (1), and hence, in the calculation of exerted force on the tip-apex in Eq. (5). To clarify this issue, note that by using Eq. (8), we expand Eq. (1) as

$$\begin{aligned} \mathbf{p}_s &= \underline{\alpha}_s^{ee} \cdot \mathbf{E}^{loc}(\mathbf{r}_s) + \hat{x}\alpha_{xx}^{em}H_x^{loc}(\mathbf{r}_s) + \hat{y}\alpha_{yy}^{em}H_y^{loc}(\mathbf{r}_s) + \hat{z}\alpha_{zz}^{em}H_z^{loc}(\mathbf{r}_s), \\ \mathbf{m}_s &= \underline{\alpha}_s^{mm} \cdot \mathbf{H}^{loc}(\mathbf{r}_s) + \hat{x}\alpha_{xx}^{me}E_x^{loc}(\mathbf{r}_s) + \hat{y}\alpha_{yy}^{me}E_y^{loc}(\mathbf{r}_s) + \hat{z}\alpha_{zz}^{me}E_z^{loc}(\mathbf{r}_s). \end{aligned} \tag{9}$$

As it is observed from this equation, because of the vanishing *z*-components of the local fields, the longitudinal components of both the magnetoelectric and electromagnetic polarizabilities $\alpha_{zz}^{me}$ and $\alpha_{zz}^{em}$ do not contribute to the determination of the induced electric and magnetic dipole moments of the sample, and hence, they do not contribute in the force calculation in Eq. (5). Indeed, based on Eq. (5), the exerted force on the tip-apex depends on the induced electric and magnetic dipole moments $\mathbf{p}_s$ and $\mathbf{m}_s$ of the sample through local fields $\mathbf{E}^{loc}(\mathbf{r}_t)$ and $\mathbf{H}^{loc}(\mathbf{r}_t)$ at the tip-apex position (see Eq. (4)). In conclusion, CP wave excitation or any excitation type that lacks the *z*-component of the electric and/or magnetic field would be an improper choice for the characterization of the longitudinal magnetoelectric polarizability component.

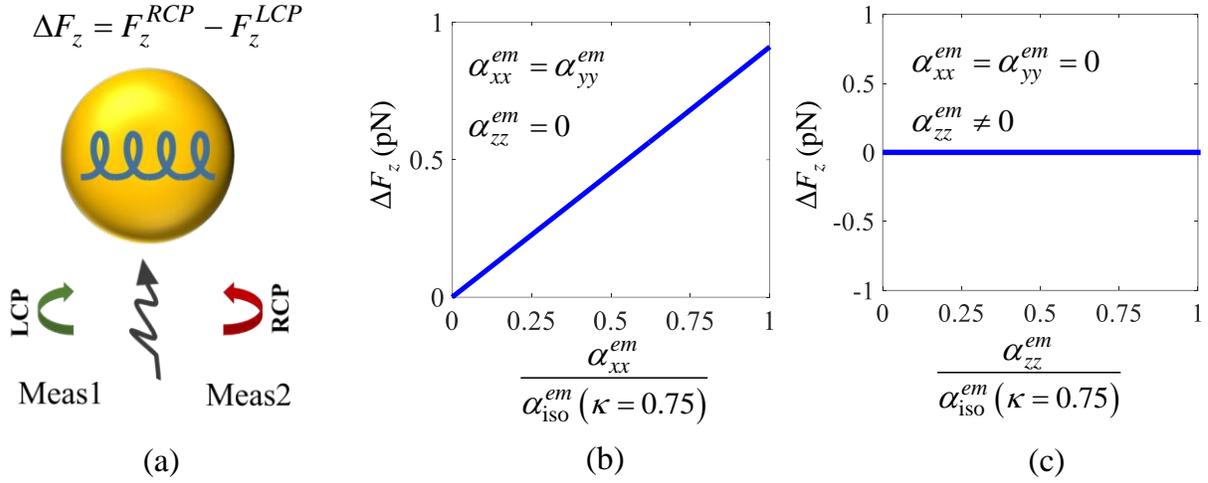

FIG. 3. (a) Schematic of appropriate structured light illumination for detection of transverse chirality. Differential force exerted on the tip-apex versus the non-vanishing component of the magnetoelectric polarizability normalized to $\alpha_{\text{iso}}^{em}\left(\kappa = 0.75\right)$ for two different cases: (b) $\alpha_{xx}^{em} = \alpha_{yy}^{em} \neq 0$ and $\alpha_{zz}^{em} = 0$; (c) $\alpha_{zz}^{em} \neq 0$ and $\alpha_{xx}^{em} = \alpha_{yy}^{em} = 0$.

Based on the above discussion, the problem of probing the longitudinal magnetoelectric polarizability component calls for utilizing alternative types of beam illuminations which is the subject of the next section.

### 4)  PROBING THE LONGITUDINAL HANDEDNESS OF CHIRAL SAMPLES

As was explained in the previous section, it is infeasible to detect the longitudinal component of the sample's magnetoelectric polarizability since the CP exciting field lacks the longitudinal field components. It is worth mentioning here that one may use CP waves which are incident obliquely with respect to the transverse plane (here, the $x$-$y$ plane), and hence, providing field components in the $z$-direction (*i.e.*, longitudinal direction in our formalism). However, utilizing a single oblique illumination results in the simultaneous presence of both the transverse and longitudinal field components and make it impossible to distinguish the transverse and longitudinal polarizability components. Furthermore, in the scenario of oblique illumination, in order to create a purely longitudinal field component we require to illuminate the tip-sample system from both sides which requires more complicated experimental set-up and special cares about the phases of exciting beams compared to our proposal.

Therefore, in order to exclusively determine the longitudinal component of the magnetoelectric polarizability it is essential to utilize illuminating beams that exclusively possess longitudinal electric and/or magnetic field components without the transverse components. Moreover, for the reason of experimental convenience, it is desirable to use beams that excite the sample from the bottom side of the surface where the sample is located, as customary in several microscopy systems. As mentioned earlier, APB and RPB are suitable candidates for our purpose since they have either purely longitudinal magnetic or electric field component along their vortex axes, respectively. Indeed, as we show next, we utilize structured light excitation made of a combination of these two beams, with proper phase difference, to retrieve the sample handedness. In other words, we use a superposition of an APB and a RPB with a specific phase difference $\psi$ (hereafter, we call it the phase parameter) with their electric and magnetic fields given by

$$E^{ARPB} = E^{APB}e^{i\psi} + E^{RPB},$$
$$H^{ARPB} = H^{APB}e^{i\psi} + H^{RPB}, \tag{10}$$

in which $E^{APB}$, $H^{APB}$, $E^{RPB}$ and $H^{RPB}$ are the electric and magnetic fields of the APB and RPB, respectively. The electric and magnetic fields of an APB in cylindrical coordinates under paraxial approximation are given by [31]

$$E^{APB} = E_\varphi \hat{\varphi} = \frac{V}{\sqrt{\pi}} \frac{2\rho}{w^2} e^{-(\rho/w)^2 \zeta} e^{-2i\tan^{-1}(z/z_R)} e^{ikz}\hat{\varphi}, \tag{11}$$

$$H^{APB} = -\frac{1}{\eta_0} E_\varphi \left[1 + \frac{1}{kz_R}\frac{\rho^2 - 2w_0^2}{w^2}\right]\hat{\rho} - \frac{V}{\sqrt{\pi}}\frac{4i}{w^2\omega\mu_0}\left[1 - \left(\frac{\rho}{w}\right)^2 \zeta\right] e^{-(\rho/w)^2 \zeta} e^{-2i\tan^{-1}(z/z_R)} e^{ikz}\hat{z}, \tag{12}$$

in which $V$ is the amplitude coefficient, $\eta_0 = \sqrt{\mu_0/\varepsilon_0}$ is the ambient wave impedance, and $z_R$ (Rayleigh range), $\zeta$ and $w$ are given by

$$w = w_0\sqrt{1 + \left(\frac{z}{z_R}\right)^2}, \quad \zeta = \left(1 - i\frac{z}{z_R}\right), \qquad z_R = \pi\frac{w_0^2}{\lambda}, \tag{13}$$

in which $w_0$ is called the beam parameter and controls the spatial extent of the beam in the transverse plane, *i.e.*, the beam waist ($w_0$ tends to be the same as the beam waist for non-sharply collimated beams). The power carried by an APB reads[31]

$$P^{APB} = \frac{|V|^2}{2\eta_0}\left(1 - \frac{1}{2(\pi w_0/\lambda)^2}\right). \tag{14}$$

For an RPB, the expressions of the electric and magnetic fields read (note that these fields are dual of the APB fields)

$$H^{RPB} = H_\varphi \hat{\varphi} = -\frac{I}{\sqrt{\pi}}\frac{2\rho}{w^2} e^{-(\rho/w)^2 \zeta} e^{-2i\tan^{-1}(z/z_R)} e^{ikz}\hat{\varphi}, \tag{15}$$

$$E^{RPB} = -\eta_0 H_\varphi \left[1 + \frac{1}{kz_R}\frac{\rho^2 - 2w_0^2}{w^2}\right]\hat{\rho} - \frac{I}{\sqrt{\pi}}\frac{4i}{w^2\omega\varepsilon_0}\left[1 - \left(\frac{\rho}{w}\right)^2 \zeta\right] e^{-(\rho/w)^2 \zeta} e^{-2i\tan^{-1}(z/z_R)} e^{ikz}\hat{z} \tag{16}$$

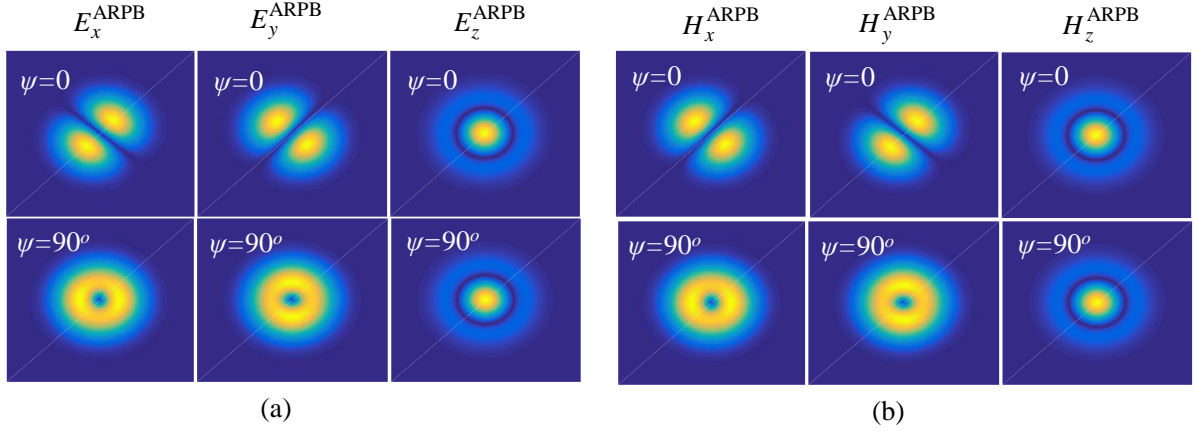

$$(a) \qquad\qquad\qquad\qquad (b)$$

FIG. 4. Magnitude of the electric (a) and magnetic (b) field components of the superposition of an APB and an RPB, for two phase parameters, $\psi = 0$ and $\psi = 90°$. On the axis, *i.e.*, at $\rho = 0$, only the $E_z$ and $H_z$ field components are non-vanishing, with $H_z^{\mathrm{ARPB}} / E_z^{\mathrm{ARPB}} = e^{i\psi} V / \left( \eta_0^2 I \right)$. In this particular case we have assumed that $V / \left( \eta_0 I \right) = 1$.

and $I$ is the amplitude coefficient of the RPB. The power carried by an RPB reads

$$P^{\mathrm{RPB}} = \eta_0 \frac{|I|^2}{2} \left( 1 - \frac{1}{2 \left( \pi w_0 / \lambda \right)^2} \right). \qquad (17)$$

On the axis of an ARPB, i.e., at $\rho = 0$, only the longitudinal field components are non-vanishing, and assuming the combined APB and RPB have the same beam parameter $w_0$ and same focal plane, the fields ratio is

$$\frac{H_z^{\mathrm{ARPB}}}{E_z^{\mathrm{ARPB}}} = \frac{H_z^{\mathrm{APB}} e^{i\psi}}{E_z^{\mathrm{RPB}}} = \frac{1}{\eta_0^2} \frac{V}{I} e^{i\psi}. \qquad (18)$$

in which $H_z^{\mathrm{ARPB}}$ and $E_z^{\mathrm{ARPB}}$ represent the $z$-component of the magnetic and electric field of the ARPB, respectively. We have demonstrated the magnitude of the components of $\mathrm{E}^{\mathrm{ARPB}}$ and $\mathrm{H}^{\mathrm{ARPB}}$ for phase parameters $\psi = 0$ and $\psi = 90°$ in Fig. 4, where we have also assumed that $F_Y \triangleq \left| \eta_0 H_z^{\mathrm{APB}} / E_z^{\mathrm{RPB}} \right| = \left| V / \left( \eta_0 I \right) \right| = 1$. We recall that the field admittance normalized to the free space wave impedance $F_Y \triangleq \left| \eta_0 H_z^{\mathrm{APB}} / E_z^{\mathrm{RPB}} \right|$ was defined previously in Refs. [31,32,49] to describe the magnetic to electric field ratio compared to that of a plane wave, since this is an important parameter when light interacts with magnetic dipoles.

As it is observed in Fig. 4, for $\psi = 0$ the magnitude of the transverse electric and magnetic field components (*i.e.*, the $x$- and $y$-components) are dumbbell-shape whereas for $\psi = 90°$ they are donut-shape. In both cases, along the beam axis (*i.e.*, along the $z$-direction, at $\rho = 0$), the transverse components of the fields vanish. Using the aforementioned combination of the beams with coincident vortex axes (the $z$-axis), assuming each of them (APB and RPB) has 1 mW power at $\lambda = 400\,\mathrm{nm}$, we place the tip-apex and chiral sample with parameters given in Fig. 2  along the axis of the two beams, on which the transverse components of the electric and magnetic fields are vanishing whereas the longitudinal components are not. Specifically, we assume the chiral sample to be on the focal plane of the beams due to high field intensity. Furthermore, we consider the beam parameter of both RPB and APB to be $w_0 = 0.7\lambda$. In order to detect the chirality, we change the phase-shift parameter $\psi$ in Eq. (10) and measure the exerted force on the tip-apex for two distinct cases: (a) a case with azimuthal chirality with $\alpha_{xx}^{em} = \alpha_{yy}^{em} = \alpha_{iso}^{em}(\kappa = 0.75)$ and $\alpha_{zz}^{em} = 0$, and b) a case with longitudinal chirality with $\alpha_{zz}^{em} = \alpha_{iso}^{em}(\kappa = 0.75)$ and $\alpha_{xx}^{em} = \alpha_{yy}^{em} = 0$, where $\alpha_{iso}^{em}(\kappa = 0.75)$ is the magnitude of magnetoelectric polarizability of the isotropic sphere of Fig. 2 with $\kappa = 0.75$. In Fig. 5 (a) and (b), we have depicted the $z$-component of the force exerted on the tip-apex versus the phase parameter $\psi$ of the ARPB. In all the following figures results are obtained including the small (though negligible) force contribution due to the magnetic polarizability of the tip-apex as in Eq. (5). As it is clearly observed, with this illumination, it is possible to distinguish between the sample's transverse and longitudinal chirality handedness by investigating the exerted force on the tip-apex by varying the values of phase parameter $\psi$. Note that the transverse component of the handedness cannot be probed by using this excitation since the exerted force on the tip-apex does not depend on the phase parameter, *i.e.*, swing of the exerted force is zero (see Fig. 5(a)).

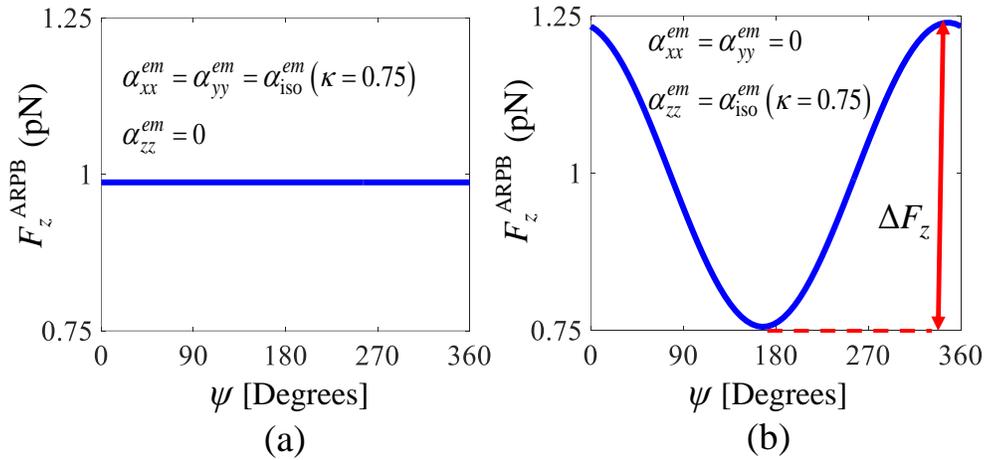

FIG. 5. Force exerted on the tip-apex in the $z$-direction when using a combination of phase-shifted APB and ARB, for two cases: (a) Case with $\alpha_{xx}^{em} = \alpha_{yy}^{em} = \alpha_{iso}^{em}(\kappa = 0.75)$  and $\alpha_{zz}^{em} = 0$ that lacks longitudinal chirality; (b) Case with $\alpha_{zz}^{em} = \alpha_{iso}^{em}(\kappa = 0.75)$ and $\alpha_{xx}^{em} = \alpha_{yy}^{em} = 0$, that has only longitudinal chirality. ARPB has $|V/(\eta_0 I)| = 1$ (*i.e.*, $F_Y = 1$ and $V$ and $I$ are in phase).

Indeed, we propose to use the "swing" of the exerted force on the tip-apex defined as

$$\Delta F_z = \max\left(F_z^{\text{ARPB}}\right) - \min\left(F_z^{\text{ARPB}}\right). \tag{19}$$

where $\max\left(F_z^{\text{ARPB}}\right)$ and $\min\left(F_z^{\text{ARPB}}\right)$ represent the maximum and the minimum of the exerted force on the tip-apex *when varying the phase shift parameter* $\psi$, observable in Fig. 5(b). These quantities are observed for two phase shift parameters that are 180 degrees apart, *i.e.* $\left|\psi_{F_z^{\max}} - \psi_{F_z^{\min}}\right| = 180^o$ as shown in Fig. 5(b) (see supplemental materials). In summary we propose to use the swing in (19) to measure longitudinal chirality. We want to quantify the physical properties that determine the force on the tip-apex, in terms of electric and magnetoelectric polarizabilities of the tip-apex and sample, respectively. In the following we provide simple formulas for the $z$-component of the force obtained from Eq. (5) by neglecting the tip-apex magnetic dipole moment, leading to $F_z^{\text{ARPB}} = \text{Re}\left[\left(\partial E_z^{\text{loc}}(z)/\partial z\right)^*_{z=z_{\text{t}}} p_{\text{t},z}\right]/2$ [45]. The force difference between two measurements for ARPB illuminations with two different phase parameters $\psi_1$ and $\psi_2$ reads (see supplemental materials)

$$F_z^{\text{ARPB}}\big|_{\psi_2} - F_z^{\text{ARPB}}\big|_{\psi_1} \approx -\frac{3F_Y\left|\text{E}_0\right|^2}{4\pi\sqrt{\varepsilon_0\mu_0}\,d^4}\Big[\text{Re}\left(e^{-i\psi_2}\alpha_{\text{t}}^{ee}\alpha_{\text{s},zz}^{em\,*}\right) - \text{Re}\left(e^{-i\psi_1}\alpha_{\text{t}}^{ee}\alpha_{\text{s},zz}^{em\,*}\right)\Big]. \quad (20)$$

Here, $\alpha_{\text{s},zz}^{em}$ is the longitudinal component of the magnetoelectric polarizability of the chiral sample and $\alpha_{\text{t}}^{ee}$ is the electric polarizability of the tip-apex. In the calculation of the above equations we have neglected the field's phase difference between the tip and sample due to their subwavelength distance. Moreover, we have simplified the electric and magnetic field expressions of the illuminating beam at the beam axis (*i.e.*, at $\rho = 0$ since both the tip-apex and sample are placed on the axis of the beam) and have rewritten Eq. (10) as

$$E^{\text{ARPB}} = E_z = \left|\text{E}_0\right|e^{-2i\tan^{-1}(z/z_R)}e^{ikz},$$
$$H^{\text{ARPB}} = H_z = \frac{\left|\text{E}_0\right|}{\eta_0}e^{-2i\tan^{-1}(z/z_R)}e^{ikz}e^{i\psi}, \quad (21)$$

where $\left|\text{E}_0\right| = 4|I|/\left(\sqrt{\pi}w^2\omega\varepsilon_0\right)$ is the amplitude of the electric field (directed along $z$) at the origin, *i.e.*, at $\rho = 0$ and $z = 0$, where the sample is located. The swing of the exerted force on the tip-apex by varying the phase parameter $\psi$ of the ARPB, neglecting the losses of the particles and assuming $F_Y = 1$ (*i.e.*, assuming that the powers of APB and RPB are equal) is given by the approximate formula (details in the supplemental materials)

$$\Delta F_z \approx -\frac{3\left|\text{E}_0\right|^2}{2\pi\sqrt{\varepsilon_0\mu_0}\,d^4}\text{Re}\left(\alpha_{\text{t}}^{ee}\right)\text{Im}\left(\alpha_{\text{s},zz}^{em}\right)^*. \quad (22)$$

In derivation of Eq. (22) we have neglected the field's phase delay in the field interaction between the tip-apex and sample, due to their subwavelength distance, *i.e.*, we have used a quasi-static Green's function[27].

Equation (22) which is similar to the differential force formula for the CP illumination, demonstrates that the sample's longitudinal magnetoelectric polarizability $\alpha_{zz}^{em}$ is determined by exploiting a combination of two vortex beams (the ARPB) and a simple algorithm where the phase shift $\psi$ between the composing APB and ARB is varied.

In Fig. 6, we further demonstrate the ability of ARPB in detecting the longitudinal chirality for two cases similar to those in Fig. 3: 1) $\alpha_{xx}^{em} = \alpha_{yy}^{em} \neq 0$ and $\alpha_{zz}^{em} = 0$; and 2) $\alpha_{zz}^{em} \neq 0$ and $\alpha_{xx}^{em} = \alpha_{yy}^{em} = 0$. We depict the swing of the $z$-component of the exerted force on the tip-apex versus the nonzero component of the magnetoelectric polarizability, normalized to the magnetoelectric polarizability of the isotropic sample sphere studied in Fig. 2 with $\kappa = 0.75$, *i.e.* $\alpha_{iso}^{em}(\kappa = 0.75)$. As one observes, this type of illumination is suitable for characterizing the longitudinal magnetoelectric polarizability $\alpha_{zz}^{em}$ since $\Delta F_z \neq 0$ in Fig. 6(c).

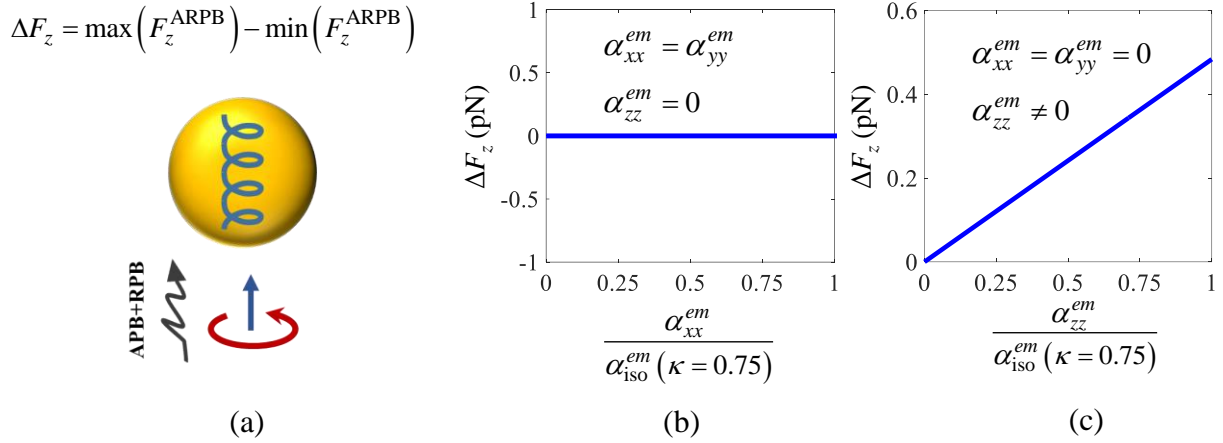

(a)    (b)    (c)

FIG. 6. *(a)* Schematic of appropriate structured light illumination for detection of *longitudinal chirality*. Maximum force swing exerted on the tip-apex versus the non-vanishing component of the magnetoelectric polarizability normalized to $\alpha_{iso}^{em}(\kappa = 0.75)$ for two different cases: *(b)* $\alpha_{xx}^{em} = \alpha_{yy}^{em} \neq 0$ and $\alpha_{zz}^{em} = 0$ ; *(c)* $\alpha_{zz}^{em} \neq 0$ and $\alpha_{xx}^{em} = \alpha_{yy}^{em} = 0$. Result in (b) demonstrates the possibility of probing the longitudinal magnetoelectric polarizability $\alpha_{zz}^{em}$ using the superposition of an APB and an RPB with proper phase difference.

Figure 7 summarizes the main results of the current study in a concise table. It demonstrates different scenarios of illumination beams and test samples. In summary, for samples with longitudinal chirality components, the ARPB is a proper choice, whereas for samples with transverse chirality components, CP beams are suitable. The suitable scenarios are highlighted by a gray color where the differential force is nonzero and the sample's chirality is retrievable from two distinct measurements.

| differential force | $F_z^{\text{RCP}} - F_z^{\text{LCP}} \neq 0$ | $F_z^{\text{RCP}} - F_z^{\text{LCP}} = 0$ | $F_z^{\text{ARPB+}} - F_z^{\text{ARPB-}} = 0$ | $F_z^{\text{ARPB+}} - F_z^{\text{ARPB-}} \neq 0$ |
|---|---|---|---|---|
| tip model | $p_{\text{t},x}$ $p_{\text{t},y}$ | $p_{\text{t},x}$ $p_{\text{t},y}$ | $p_{\text{t},z}$ | $p_{\text{t},z}$ |
| sample model | $p_{\text{s},x}$ $m_{\text{s},x}$ $p_{\text{s},y}$ $m_{\text{s},y}$ | $p_{\text{s},z}$ $m_{\text{s},z}$ | $p_{\text{s},x}$ $m_{\text{s},x}$ $p_{\text{s},y}$ $m_{\text{s},y}$ | $p_{\text{s},z}$ $m_{\text{s},z}$ |
| illumination | LCP Meas1  Meas2 RCP | LCP Meas1  Meas2 RCP | ARPB- Meas1  Meas2 ARPB+ | ARPB- Meas1  Meas2 ARPB+ |

FIG. 7. Summary of proper illumination type to retrieve chirality of anisotropic chiral samples. In each case two measurements Meas1 and Meas2 are necessary. An experiment with two CP beams is (not) able to determine the transverse (longitudinal) chirality. An experiment with two ARPB is (not) able to determine the longitudinal (transverse) chirality. ARPB$^+$ and ARPB$^{--}$ represent the two APRBs which lead to $\max\left(F_z^{\text{ARPB}}\right)$ and $\min\left(F_z^{\text{ARPB}}\right)$, respectively. The chirality in each direction is schematized by a helix and the proper choice of excitation is highlighted with a shaded gray area. In all cases the tip-apex is modeled by an electric polarizable dipole and the choice of the excitation influences the direction of the induced dipole on the tip-apex although the tip-apex is isotropic.

In the previous discussion and results we have assumed that $\eta_0 H_z^{\text{APB}} / E_z^{\text{RPB}} = V / (\eta_0 I) = 1$, which means that both the APB and RPB carry the same power (assuming they have the same beam waist) and $V$ and $I$ have the same phase, since the phase difference between these two beams is represented by $\psi$.

Our final discussion before conclusion is about having different illumination powers of the APB and RPB, and then we justify the choice done beforehand, where $F_Y = 1$, that leads to the best result. Therefore we now assume that the parameter $F_Y^2$ differs from unity, where $F_Y^2$ is the figure of merit which shows the ratio between the illumination power of the APB and RPB, previously defined as

$$F_Y^2 = \left|\frac{P^{\text{APB}}}{P^{\text{RPB}}}\right| = \eta_0^2 \left|\frac{H_z}{E_z}\right|^2_{\substack{\rho=0 \\ z=0}} = \frac{1}{\eta_0^2}\left|\frac{V}{I}\right|^2 . \tag{23}$$

Note that $F_Y^2$ represents the intensity of the longitudinal magnetic field over the intensity of the longitudinal electric field, at the focus of the beams (*i.e.*, at the chiral sample location). This figure of merit is the squared magnitude of the normalized field admittance introduced in previous studies [31,32,49]. We assume that the *total* power of the APB and RPB exciting beams is constant, and in particular $P^{\text{APB}} + P^{\text{RPB}} = 2\,\text{mW}$ (there is no cross power term since the polarization of the two beams is orthogonal). We also assume that the tip-apex and the sample have the same polarizability characteristics used in the example in Fig. 5 (a), *i.e.*, $\alpha_{zz}^{em} = \alpha_{iso}^{em}(\kappa = 0.75)$ and $\alpha_{xx}^{em} = \alpha_{yy}^{em} = 0$. The color map in Fig. 8 (a) describes the exerted force on the tip-apex versus $F_Y^2$ (in logarithmic scale) and phase difference $\psi$ between the APB and RPB. One observes that as $F_Y^2$ increases (increases the ratio between the APB to RPB power, or alternatively the magnetic field increases with respect to the electric field), the exerted force on the tip-apex decreases, for all phase parameters $\psi$. The reason is that for both the tip-apex and sample, in general the electric responses are stronger than their magnetic counterparts, thus, weaker electric field (larger $F_Y^2$) leads to weaker observed optical force on the tip-apex. However, it is important to note that for a reliable measurement, the most important property is to observe the *largest swing of the force i.e.* $\max\left(\Delta F_z\right)$ (note that $\Delta F_z$ is defined in Eq. (19)) which leads to higher resolution in detection of nanoscale chirality, and this feature may be more interesting than a large observable force. In other words, a stronger force does not necessarily lead to a larger swing force. In Fig. 8(b) we have depicted $\Delta F_z$ versus $F_Y^2$, and we observe that the maximum swing occurs when $F_Y^2 = 1$, *i.e.*, when the illumination powers of the APB and RPB are equal or equivalently when $|E_z|/|H_z| = \eta_0$ or $|V|/|I| = \eta_0$. Indeed, it can be analytically proved (details in the supplemental materials) that assuming a constant total power of APB and RPB (*i.e.*, a constant power density), the maximum swing occurs when $F_Y = 1$.

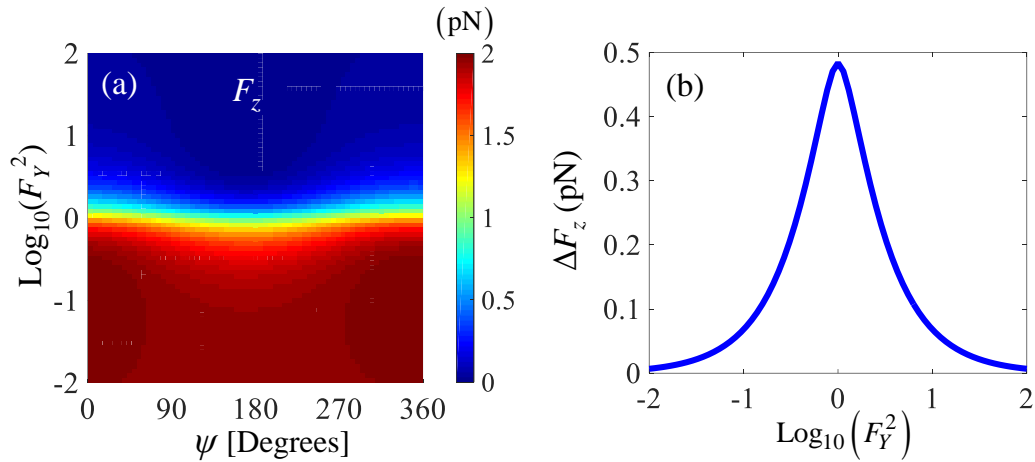

FIG. 8. (a) Photo-induced optical force on the tip-apex varying the power distribution between the APB and the RPB, and their phase difference $\psi$. The force increases with decreasing $F_Y^2$. (b) Swing of the force versus $F_Y^2$. In conclusion the maximum swing corresponds to the best possible scenario for longitudinal chirality detection, and it occurs when $F_Y^2 = 1$, or equivalently when $|E_z|/|H_z| = \eta_0$.

With the above considerations, we conclude the discussion on introducing suitable high resolution approaches to unscramble the chirality structure of a chiral sample. However, future work is necessary to distinguish between the two transverse components of the magnetoelectric polarizability tensor, *i.e.*, $\alpha_{xx}^{em}$ and $\alpha_{yy}^{em}$. For that azimuthally anisotropic case, excitation types that possess either a *pure* electric or a *pure* magnetic field component in the transverse plane *i.e.*, the $xy$-plane, would be useful following the scheme proposed in this paper.

### 5) CONCLUSION

Based on the concept of photo-induced optical force, we have presented how different structured-light excitation scenarios result in chirality characterization at nanoscale. We have shown an approach to determine not only the enantiomer type of a chiral sample but also to distinguish between its transverse and longitudinal chirality components at nanoscale. Particularly, we have proved that when we are limited to the illumination from the bottom side of the tip-sample system (as shown in Fig. 1), probing the transverse sample's chirality requires transverse field components (with respect to the propagation direction) as in CP beams. Instead, probing longitudinal chirality is achievable by using a combination of longitudinal field components, obtained by performing two experiments, each one with a superposition of an APB and an RPB with proper phase shift $\psi$. There is an underlined difference between CP beams, whose light is chiral, and APBs and RPBs, whose light is not chiral. However combinations of an APB and an RPB leads to chiral light, therefore we propose to use them together with a proper controlled phased shift $\psi$, and perform two experiments, with two different phase shifts. The difference between these two experiments, denoted as $\Delta F_z$, takes the maximum value when the APB and RPB exhibit the same power density at the chiral sample location. The quantification of the longitudinal and transverse components of the magnetoelectric polarizability helps to unscramble the structure of chiral specimen. This work has the potential to advance studies of chirality of nanosamples and molecular concentrations, since chirality is a fundamental building block of nature.

### ACKNOWLEDGEMENT

The authors acknowledge support from the W. M. Keck Foundation, USA.

### REFERENCES

(1)    Ambilino, D. B. *Chirality at the Nanoscale: Nanoparticles, Surfaces, Materials and More*; Wiley-VCH: Germany, 2009.

(2)    Freedman, T. B.; Cao, X.; Dukor, R. K.; Nafie, L. A. Absolute Configuration Determination of Chiral Molecules in the Solution State Using Vibrational Circular Dichroism. *Chirality* **2003**, *15* (9), 743–758.

(3)    Barron, L. D. An Introduction to Chirality at the Nanoscale. In *Chirality at the Nanoscale*; Amabilino, D. B., Ed.; Wiley-VCH Verlag GmbH & Co. KGaA, 2009; pp 1–27.

(4)    Lodish, H.; Berk, A.; Zipursky, S. L.; Matsudaira, P.; Baltimore, D.; Darnell, J. *Molecular Cell Biology*, 4th ed.; W. H. Freeman, 2000.

(5)    Heim, M. A.; Jakoby, M.; Werber, M.; Martin, C.; Weisshaar, B.; Bailey, P. C. The Basic Helix–Loop–Helix Transcription Factor Family in Plants: A Genome-Wide Study of Protein Structure and Functional Diversity. *Mol. Biol. Evol.* **2003**, *20* (5), 735–747.

(6)    Joye, I. J.; Lagrain, B.; Delcour, J. A. Use of Chemical Redox Agents and Exogenous Enzymes to Modify the Protein Network during Breadmaking – A Review. *J. Cereal Sci.* **2009**, *50* (1), 11–21.

(7)    Mortimer, S. A.; Kidwell, M. A.; Doudna, J. A. Insights into RNA Structure and Function from Genome-Wide Studies. *Nat. Rev. Genet.* **2014**, *15* (7), 469–479.

(8)    Moult, J.; Fidelis, K.; Kryshtafovych, A.; Schwede, T.; Tramontano, A. Critical Assessment of Methods of Protein Structure Prediction: Progress and New Directions in Round XI. *Proteins Struct. Funct. Bioinforma.* **2016**, *84*, 4–14.

(9)     Huang, Y.; Liu, L. Protein Modification: Standing out from the Crowd. *Nat. Chem.* **2016**, *8* (2), 101–102.

(10)    Barron, L. D.; Buckingham, A. D. Rayleigh and Raman Scattering from Optically Active Molecules. *Mol. Phys.* **1971**, *20* (6), 1111–1119.

(11)    Hug, W.; Surbeck, H. Vibrational Raman Optical Activity Spectra Recorded in Perpendicular Polarization. *Chem. Phys. Lett.* **1979**, *60* (2), 186–192.

(12)    Barron, L. D. Raman Optical Activity. In *Optical Activity and Chiral Discrimination*; Mason, S. F., Ed.; NATO Advanced Study Institutes Series; Springer Netherlands, 1979; pp 219–262.

(13)    Barron, L. D.; Blanch, E. W.; Hecht, L. Unfolded Proteins Studied by Raman Optical Activity; Chemistry, B.-A. in P., Ed.; Unfolded Proteins; Academic Press, 2002; Vol. 62, pp 51–90.

(14)    Johnson, W. C. Protein Secondary Structure and Circular Dichroism: A Practical Guide. *Proteins Struct. Funct. Bioinforma.* **1990**, *7* (3), 205–214.

(15)    Kelly, S. M.; Jess, T. J.; Price, N. C. How to Study Proteins by Circular Dichroism. *Biochim. Biophys. Acta BBA - Proteins Proteomics* **2005**, *1751* (2), 119–139.

(16)    Greenfield, N. J. Using Circular Dichroism Spectra to Estimate Protein Secondary Structure. *Nat. Protoc.* **2007**, *1* (6), 2876–2890.

(17)    Martin, Y.; Williams, C. C.; Wickramasinghe, H. K. Atomic Force Microscope–force Mapping and Profiling on a Sub 100-Å Scale. *J. Appl. Phys.* **1987**, *61* (10), 4723–4729.

(18)    Martin, Y.; Abraham, D. W.; Wickramasinghe, H. K. High-resolution Capacitance Measurement and Potentiometry by Force Microscopy. *Appl. Phys. Lett.* **1988**, *52* (13), 1103–1105.

(19)    Rajapaksa, I.; Uenal, K.; Wickramasinghe, H. K. Image Force Microscopy of Molecular Resonance: A Microscope Principle. *Appl. Phys. Lett.* **2010**, *97* (7), 073121.

(20)    Rajapaksa, I.; Wickramasinghe, H. K. Raman Spectroscopy and Microscopy Based on Mechanical Force Detection. *Appl. Phys. Lett.* **2011**, *99* (16), 161103.

(21)    Jahng, J.; Brocious, J.; Fishman, D. A.; Huang, F.; Li, X.; Tamma, V. A.; Wickramasinghe, H. K.; Potma, E. O. Gradient and Scattering Forces in Photoinduced Force Microscopy. *Phys. Rev. B* **2014**, *90* (15), 155417.

(22)    Guclu, C.; Tamma, V. A.; Wickramasinghe, H. K.; Capolino, F. Photo-Induced Magnetic Force Between Nanostructures. *Phys. Rev. B* **2015**, *92* (23).

(23)    Jahng, J.; Fishman, D. A.; Park, S.; Nowak, D. B.; Morrison, W. A.; Wickramasinghe, H. K.; Potma, E. O. Linear and Nonlinear Optical Spectroscopy at the Nanoscale with Photoinduced Force Microscopy. *Acc. Chem. Res.* **2015**, *48* (10), 2671–2679.

(24)    Huang, F.; Tamma, V. A.; Rajaei, M.; Almajhadi, M.; Kumar Wickramasinghe, H. Measurement of Laterally Induced Optical Forces at the Nanoscale. *Appl. Phys. Lett.* **2017**, *110* (6), 063103.

(25)    Rajaei, M.; Almajhadi, M. A.; Zeng, J.; Wickramasinghe, H. K. Near-Field Nanoprobing Using Si Tip-Au Nanoparticle Photoinduced Force Microscopy with 120:1 Signal-to-Noise Ratio, Sub-6-Nm Resolution. *ArXiv180405993 Phys.* **2018**.

(26)    Zeng, J.; Huang, F.; Guclu, C.; Veysi, M.; Albooyeh, M.; Wickramasinghe, H. K.; Capolino, F. Sharply Focused Azimuthally Polarized Beams with Magnetic Dominance: Near-Field Characterization at Nanoscale by Photoinduced Force Microscopy. *ACS Photonics* **2018**, *5* (2), 390–397.

(27)   Kamandi, M.; Albooyeh, M.; Guclu, C.; Veysi, M.; Zeng, J.; Wickramasinghe, K.; Capolino, F. Enantiospecific Detection of Chiral Nanosamples Using Photoinduced Force. *Phys. Rev. Appl.* **2017**, *8* (6), 064010.

(28)   Zhao, Y.; Saleh, A. A. E.; Haar, M. A. van de; Baum, B.; Briggs, J. A.; Lay, A.; Reyes-Becerra, O. A.; Dionne, J. A. Nanoscopic Control and Quantification of Enantioselective Optical Forces. *Nat. Nanotechnol.* **2017**, *12* (11), 1055–1059.

(29)   Coullet, P.; Gil, L.; Rocca, F. Optical Vortices. *Opt. Commun.* **1989**, *73* (5), 403–408.

(30)   Veysi, M.; Guclu, C.; Capolino, F. Vortex Beams with Strong Longitudinally Polarized Magnetic Field and Their Generation by Using Metasurfaces. *JOSA B* **2015**, *32* (2), 345–354.

(31)   Veysi, M.; Guclu, C.; Capolino, F. Focused Azimuthally Polarized Vector Beam and Spatial Magnetic Resolution below the Diffraction Limit. *JOSA B* **2016**, *33* (11), 2265–2277.

(32)   Guclu, C.; Veysi, M.; Capolino, F. Photoinduced Magnetic Nanoprobe Excited by an Azimuthally Polarized Vector Beam. *ACS Photonics* **2016**, *3* (11), 2049–2058.

(33)   Nowak, D.; Morrison, W.; Wickramasinghe, H. K.; Jahng, J.; Potma, E.; Wan, L.; Ruiz, R.; Albrecht, T. R.; Schmidt, K.; Frommer, J.; et al. Nanoscale Chemical Imaging by Photoinduced Force Microscopy. *Sci. Adv.* **2016**, *2* (3), e1501571.

(34)   Serdyukov, A.; Semchenko, I.; Tretyakov, S.; Sihvola, A. *Electromagnetics of Bi-Anisotropic Materials : Theory and Applications*; Gordon and Breach Science: Australia, 2001.

(35)   Campione, S.; Capolino, F. Ewald Method for 3D Periodic Dyadic Green's Functions and Complex Modes in Composite Materials Made of Spherical Particles under the Dual Dipole Approximation. *Radio Sci.* **2012**, *47* (6), RS0N06.

(36)   Rahimzadegan, A.; Fruhnert, M.; Alaee, R.; Fernandez-Corbaton, I.; Rockstuhl, C. Optical Force and Torque on Dipolar Dual Chiral Particles. *Phys. Rev. B* **2016**, *94* (12), 125123.

(37)   Varault, S.; Rolly, B.; Boudarham, G.; Demésy, G.; Stout, B.; Bonod, N. Multipolar Effects on the Dipolar Polarizability of Magneto-Electric Antennas. *Opt. Express* **2013**, *21* (14), 16444–16454.

(38)   Jackson, D. *Classical Electrodynamics*; Wiley: New York, 1998.

(39)   Kamandi, M.; Emadi, S.-M.-H.; Faraji-Dana, R. Integral Equation Analysis of Multilayered Waveguide Bends Using Complex Images Green's Function Technique. *J. Light. Technol.* **2015**, *33* (9), 1774–1779.

(40)   Yaghjian, A. D. Electromagnetic Forces on Point Dipoles. In *IEEE Antennas and Propagation Society International Symposium*; Orlando, FL, USA, 1999; Vol. 4, pp 2868–2871.

(41)   Chaumet, P. C.; Nieto-Vesperinas, M. Time-Averaged Total Force on a Dipolar Sphere in an Electromagnetic Field. *Opt. Lett.* **2000**, *25* (15), 1065–1067.

(42)   Nieto-Vesperinas, M.; Sáenz, J. J.; Gómez-Medina, R.; Chantada, L. Optical Forces on Small Magnetodielectric Particles. *Opt. Express* **2010**, *18* (11), 11428–11443.

(43)   Wang, S. B.; Chan, C. T. Lateral Optical Force on Chiral Particles near a Surface. *Nat. Commun.* **2014**, *5*, 3307.

(44)   Alaee, R.; Gurlek, B.; Christensen, J.; Kadic, M. Optical Force Rectifiers Based on PT-Symmetric Metasurfaces. *Phys. Rev. B* **2018**, *97* (19), 195420.

(45)   Gordon, J. P. Radiation Forces and Momenta in Dielectric Media. *Phys. Rev. A* **1973**, *8* (1), 14–21.

(46)    Gao, D.; Ding, W.; Nieto-Vesperinas, M.; Ding, X.; Rahman, M.; Zhang, T.; Lim, C.; Qiu, C.-W. Optical Manipulation from the Microscale to the Nanoscale: Fundamentals, Advances and Prospects. *Light Sci. Appl.* **2017**, *6* (9), e17039.

(47)    Albooyeh, M.; Hanifeh, M.; Kamandi, M.; Rajaei, M.; Zeng, J.; Wickramasinghe, H. K.; Capolino, F. Photo-Induced Force vs Power in Chiral Scatterrers. In *2017 IEEE International Symposium on Antennas and Propagation USNC/URSI National Radio Science Meeting*; 2017; pp 35–36.

(48)    Zeng, J.; Kamandi, M.; Darvishzadeh-Varcheie, M.; Albooyeh, M.; Veysi, M.; Guclu, C.; Hanifeh, M.; Rajaei, M.; Potma, E. O.; Wickramasinghe, H. K.; et al. In Pursuit of Photo-Induced Magnetic and Chiral Microscopy. *EPJ Appl. Metamaterials* **2018**, *5*, 7.

(49)    Darvishzadeh-Varcheie, M.; Guclu, C.; Capolino, F. Magnetic Nanoantennas Made of Plasmonic Nanoclusters for Photoinduced Magnetic Field Enhancement. *Phys. Rev. Appl.* **2017**, *8* (2), 024033.

(50)    Darvishzadeh-Varcheie, M.; Guclu, C.; Ragan, R.; Boyraz, O.; Capolino, F. Electric Field Enhancement with Plasmonic Colloidal Nanoantennas Excited by a Silicon Nitride Waveguide. *Opt. Express* **2016**, *24* (25), 28337–28352.

(51)    Sihvola, A. H.; Lindell, I. V. Chiral Maxwell-Garnett Mixing Formula. *Electron. Lett.* **1990**, *26* (2), 118–119.

(52)    Zhao, Y.; Saleh, A. A. E.; Dionne, J. A. Enantioselective Optical Trapping of Chiral Nanoparticles with Plasmonic Tweezers. *ACS Photonics* **2016**, *3* (3), 304–309.

(53)    Mohammadi-Baghaee, R.; Rashed-Mohassel, J. The Chirality Parameter for Chiral Chemical Solutions. *J. Solut. Chem.* **2016**, *45* (8), 1171–1181.

(54)    Penzkofer, A. Optical Rotatory Dispersion Measurement of D-Glucose with Fixed Polarizer Analyzer Accessory in Conventional Spectrophotometer. *J. Anal. Sci. Instrum.* **2013**, *Vol. 3*, 234–239.

(55)    Lambert, J.; Compton, R. N.; Crawford, T. D. The Optical Activity of Carvone: A Theoretical and Experimental Investigation. *J. Chem. Phys.* **2012**, *136* (11), 114512.

(56)    Bruice, P. Y. *Organic Chemistry*, 6th ed.; Pearson: United States, 2011.

(57)    Katz, T. J.; Nuckolls, C. P. Aggregates of Substituted (6)Helicene Compounds That Show Enhanced Optical Rotatory Power and Nonlinear Optical Response and Uses Thereof. US5993700 A, November 30, 1999.

(58)    Caricato, M.; Vaccaro, P. H.; Crawford, T. D.; Wiberg, K. B.; Lahiri, P. Insights on the Origin of the Unusually Large Specific Rotation of (1S,4S)-Norbornenone. *J. Phys. Chem. A* **2014**, *118* (26), 4863–4871.

(59)    Yang, J.; Somlyo, A. P.; Siao, Z.; Mou, J. Cryogenic Atomic Force Microscope. US5410910 A, May 2, 1995.

(60)    Hansen, P. M.; Bhatia, V. K.; Harrit, N.; Oddershede, L. Expanding the Optical Trapping Range of Gold Nanoparticles. *Nano Lett.* **2005**, *5* (10), 1937–1942.

(61)    Selhuber-Unkel, C.; Zins, I.; Schubert, O.; Sönnichsen, C.; Oddershede, L. B. Quantitative Optical Trapping of Single Gold Nanorods. *Nano Lett.* **2008**, *8* (9), 2998–3003.

(62)    Hajizadeh, F.; S.Reihani, S. N. Optimized Optical Trapping of Gold Nanoparticles. *Opt. Express* **2010**, *18* (2), 551–559.

(63)    Bohren, C. F. PhD Thesis, University of Arizona, 1975.

**Supplemental Materials for article**

**"Unscrambling Structured Chirality with Structured Light at Nanoscale Using Photo-induced Force"**

Mohammad Kamandi, Mohammad Albooyeh, Mehdi Veysi, Mohsen Rajaei, Jinwei Zeng,

Kumar Wickramasinghe and Filippo Capolino

Department of Electrical Engineering and Computer Science, University of California, Irvine, California 92697, USA

**1)  LOCAL FIELDS AT THE TIP-APEX AND SAMPLE LOCATION, EXCITED BY AN ARPB**

We prove that under ARPB excitation (a superposition of two coaxial beams: an APB and an RPB with proper phase shift $\psi$ ) the local fields at the tip-apex and sample locations (both on the ARPB axis, see Fig. 1 of the paper) lack transverse components and we determine the longitudinal field components that include the near-field interaction. We consider the schematic of the problem in Fig. 1 of the manuscript and assume that the tip-apex and chiral sample are located at $z_\mathrm{t}$ and $z_\mathrm{s}$, respectively, at a distance $d$ from each other. Here subscripts "s" and "t" denote sample and tip, respectively. According to Eqs. (B6) and (B7) in Ref. [1] the local electric and magnetic fields at the position of the tip-apex are

$$\begin{bmatrix} E_x^{\mathrm{loc}}(z_\mathrm{t}) \\ E_y^{\mathrm{loc}}(z_\mathrm{t}) \\ E_z^{\mathrm{loc}}(z_\mathrm{t}) \end{bmatrix} = \begin{bmatrix} E_x^{\mathrm{inc}}(z_\mathrm{t}) \\ E_y^{\mathrm{inc}}(z_\mathrm{t}) \\ E_z^{\mathrm{inc}}(z_\mathrm{t}) \end{bmatrix} - \frac{G}{\varepsilon_0} \begin{bmatrix} \alpha_\mathrm{s}^{ee} E_x^{\mathrm{loc}}(z_\mathrm{s}) + \alpha_\mathrm{s}^{em} H_x^{\mathrm{loc}}(z_\mathrm{s}) \\ \alpha_\mathrm{s}^{ee} E_y^{\mathrm{loc}}(z_\mathrm{s}) + \alpha_\mathrm{s}^{em} H_y^{\mathrm{loc}}(z_\mathrm{s}) \\ -2\alpha_\mathrm{s}^{ee} E_z^{\mathrm{loc}}(z_\mathrm{s}) - 2\alpha_\mathrm{s}^{em} H_z^{\mathrm{loc}}(z_\mathrm{s}) \end{bmatrix}, \tag{1}$$

and

$$\begin{bmatrix} H_x^{\mathrm{loc}}(z_\mathrm{t}) \\ H_y^{\mathrm{loc}}(z_\mathrm{t}) \\ H_z^{\mathrm{loc}}(z_\mathrm{t}) \end{bmatrix} = \begin{bmatrix} H_x^{\mathrm{inc}}(z_\mathrm{t}) \\ H_y^{\mathrm{inc}}(z_\mathrm{t}) \\ H_z^{\mathrm{inc}}(z_\mathrm{t}) \end{bmatrix} - \frac{G}{\mu_0} \begin{bmatrix} \alpha_\mathrm{s}^{me} E_x^{\mathrm{loc}}(z_\mathrm{s}) + \alpha_\mathrm{s}^{mm} H_x^{\mathrm{loc}}(z_\mathrm{s}) \\ \alpha_\mathrm{s}^{me} E_y^{\mathrm{loc}}(z_\mathrm{s}) + \alpha_\mathrm{s}^{mm} H_y^{\mathrm{loc}}(z_\mathrm{s}) \\ -2\alpha_\mathrm{s}^{me} E_z^{\mathrm{loc}}(z_\mathrm{s}) - 2\alpha_\mathrm{s}^{mm} H_z^{\mathrm{loc}}(z_\mathrm{s}) \end{bmatrix}, \tag{2}$$

respectively. Because of the subwavelength distance between the tip-apex and the chiral sample, the free space Green's function is approximated by its strongest term (see Ref [1])

$$G = \frac{e^{ik|z_\mathrm{t}-z_\mathrm{s}|}}{4\pi|z_\mathrm{t}-z_\mathrm{s}|^3}. \tag{3}$$

Combining Eqs. (B8)-(B10) in Ref [1] and noting that the incident azimuthally and radially polarized beams lack transverse components on their axis (the $z$ axis), and since both the tip-apex and sample are located along the optical axis, Eqs. (1) and (2) reduce to

$$
\begin{bmatrix} E_x^{\text{loc}}(z_{\text{t}}) \\[6pt] E_y^{\text{loc}}(z_{\text{t}}) \\[6pt] E_z^{\text{loc}}(z_{\text{t}}) \end{bmatrix} = \begin{bmatrix} 0 \\[6pt] 0 \\[6pt] E_z^{\text{inc}}(z_{\text{t}}) \end{bmatrix} - \frac{G}{\varepsilon_0} \begin{bmatrix} \alpha_{\text{s}}^{ee}\left(-\dfrac{G}{\varepsilon_0}\alpha_{\text{t}}^{ee}E_x^{\text{loc}}(z_{\text{t}})\right) \\[10pt] \alpha_{\text{s}}^{ee}\left(-\dfrac{G}{\varepsilon_0}\alpha_{\text{t}}^{ee}E_y^{\text{loc}}(z_{\text{t}})\right) \\[10pt] -2\alpha_{\text{s}}^{ee}\left(E_z^{\text{inc}}(z_{\text{t}})+2\dfrac{G}{\varepsilon_0}\alpha_{\text{t}}^{ee}E_z^{\text{loc}}(z_{\text{t}})\right) - 2\alpha_{\text{s}}^{em}H_z^{\text{inc}}(z_{\text{t}}) \end{bmatrix}, \quad (4)
$$

and

$$
\begin{bmatrix} H_x^{\text{loc}}(z_{\text{t}}) \\[6pt] H_y^{\text{loc}}(z_{\text{t}}) \\[6pt] H_z^{\text{loc}}(z_{\text{t}}) \end{bmatrix} = \begin{bmatrix} 0 \\[6pt] 0 \\[6pt] H_z^{\text{inc}}(z_{\text{t}}) \end{bmatrix} - \frac{G}{\mu_0} \begin{bmatrix} \alpha_{s}^{me}\left(-\dfrac{G}{\varepsilon_0}\alpha_{\text{t}}^{ee}E_x^{\text{loc}}(z_{\text{t}})\right) \\[10pt] \alpha_{s}^{me}\left(-\dfrac{G}{\varepsilon_0}\alpha_{\text{t}}^{ee}E_y^{\text{loc}}(z_{\text{t}})\right) \\[10pt] -2\alpha_{\text{s}}^{me}\left(E_z^{\text{inc}}(z_{\text{t}})+2\dfrac{G}{\varepsilon_0}\alpha_{\text{t}}^{ee}E_z^{\text{loc}}(z_{\text{t}})\right) - 2\alpha_{\text{s}}^{mm}H_z^{\text{inc}}(z_{\text{t}}) \end{bmatrix}, \quad (5)
$$

respectively. Solving Eqs. (4) and (5) for the longitudinal field components (there are 6 equations and 6 unknowns), the transverse $x$ and $y$ components of the local fields vanish and the longitudinal $z$-component of the electric and magnetic fields read

$$
E_z^{\text{loc}}(z_{\text{t}}) = \frac{1}{1-\left(2G\right)^2 \dfrac{\alpha_{\text{t}}^{ee}}{\varepsilon_0}\dfrac{\alpha_{\text{s}}^{ee}}{\varepsilon_0}}\left[\left(1+2G\frac{\alpha_{\text{s}}^{ee}}{\varepsilon_0}\right)E_z^{\text{inc}}(z_{\text{t}})+2G\frac{\alpha_{\text{s}}^{em}}{\varepsilon_0}H_z^{\text{inc}}(z_{\text{t}})\right], \quad (6)
$$

and

$$
H_z^{\text{loc}}(z_{\text{t}}) = \left(2G\frac{\alpha_{s}^{me}}{\mu_0}+\frac{\left(1+2G\dfrac{\alpha_{\text{s}}^{ee}}{\varepsilon_0}\right)\left(2G\right)^2}{1-\left(2G\right)^2\dfrac{\alpha_{\text{t}}^{ee}}{\varepsilon_0}\dfrac{\alpha_{\text{s}}^{ee}}{\varepsilon_0}}\frac{\alpha_{\text{t}}^{ee}}{\varepsilon_0}\frac{\alpha_{s}^{me}}{\mu_0}\right)E_z^{\text{inc}}(z_{\text{t}})+
$$

$$
\left(1+2G\frac{\alpha_{\text{s}}^{mm}}{\mu_0}+\frac{\left(2G\right)^3}{1-\left(2G\right)^2\dfrac{\alpha_{\text{t}}^{ee}}{\varepsilon_0}\dfrac{\alpha_{s}^{ee}}{\varepsilon_0}}\frac{\alpha_{\text{t}}^{ee}}{\varepsilon_0}\frac{\alpha_{s}^{me}}{\mu_0}\frac{\alpha_{s}^{em}}{\varepsilon_0}\right)H_z^{\text{inc}}(z_{\text{t}}), \quad (7)
$$

respectively, where we have neglected the phase difference (i.e., the retardation effect) between the tip-apex and sample since their distance is electrically small, and we have chosen to have slowly varying incident field with $z$ compared to the scattered near-fields [2]. Using a similar approach, one may derive the local electric and magnetic fields at the chiral sample location, and it is also clear that for ARPB excitation, the local electric and magnetic fields at both the tip-apex and sample locations lack transverse components.

## 2) SWING OF THE OPTICAL FORCE UNDER ARPB EXCITATION

In this section we summarize the steps to obtain an approximate formulation for the *swing* of the force exerted on the tip-apex in the near-field region of a chiral sample under ARPB illumination discussed in Sec. (4) of the manuscript. From Eq. (5) of the manuscript, after neglecting the effect of magnetic dipole of the tip-apex (compared to its electric dipole effect), the expression for the time-averaged optical force exerted on the achiral tip-apex, represented solely as an electric dipole $p_t$, reads

$$F = \frac{1}{2} \text{Re}\left[ \left( \nabla E^{\text{loc}}(\mathbf{r}) \right)^*_{\mathbf{r}=\mathbf{r}_t} \cdot p_t \right]. \tag{8}$$

The gradient operator for vector $\mathbf{A}$, in matrix representation based on Cartesian coordinates is defined as

$$\nabla A = \begin{bmatrix} \partial_x A_x & \partial_y A_x & \partial_z A_x \\ \partial_x A_y & \partial_y A_y & \partial_z A_y \\ \partial_x A_z & \partial_y A_z & \partial_z A_z \end{bmatrix}. \tag{9}$$

Since local APRB fields possess only $z$-components on the beam axis, the tip-apex electric dipole has only the $z$-component, and as a consequence the force exerted on the tip-apex possesses only the $z$-component, given by

$$F_z^{\text{ARPB}} = \frac{1}{2} \text{Re}\left[ \left( \frac{\partial}{\partial z} E_z^{\text{loc}}(z) \right)^*_{z=z_t} p_{t,z} \right]. \tag{10}$$

Noting that $E_z^{\text{loc}}(z_t) = E_z^{\text{inc}}(z_t) + 2 p_{s,z} G / \varepsilon_0$, and assuming that $E_z^{\text{inc}}$ is slowly varying with $z$ compared to the scattered field generated by the sample, we obtain[2]

$$F_z^{\text{ARPB}} \approx \frac{1}{2} \text{Re}\left[ \left( 2 p_{s,z} \frac{\partial}{\partial z} \frac{G}{\varepsilon_0} \right)^*_{z=z_t} p_{t,z} \right]. \tag{11}$$

We consider the ARPB incident fields, that on the beam axis (i.e., at $\rho = 0$ since both the tip and sample are placed on the axis of the beam) reduce to

$$E_z^{\text{ARPB}} = E_z^{\text{RPB}} = |\text{E}_0| e^{-2i \tan^{-1}(z/z_R)} e^{ikz},$$
$$H_z^{\text{ARPB}} = H_z^{\text{APB}} e^{i\varphi} = F_Y \frac{|\text{E}_0|}{\eta_0} e^{-2i \tan^{-1}(z/z_R)} e^{ikz} e^{i\psi}. \tag{12}$$

where $\psi$ is the phase shift between the APB and the RPB constituents, as defined in Eq. (10) of the manuscript and $|\text{E}_0| = 4 |I| / \left( \sqrt{\pi} w^2 \omega \varepsilon_0 \right)$ is the amplitude of the electric field at the origin , i.e., at $\rho = 0$ and $z = 0$.

Note that when $F_Y = 1$, the powers of the APB and RPB illuminations (assumed to have the same beam waist) are equal. Next, considering the quasistatic limit by assuming $kd \to 0$, where $d$ is the distance between the sample and the tip (Fig. 1 of the manuscript), and neglecting the terms that contain polarizability power orders higher than the second, Eq. (11) leads to

$$F_z^{\text{ARPB}} \approx \frac{1}{2} \text{Re}\left[\left(\frac{\alpha_s^{ee}}{\varepsilon_0} + F_Y \frac{\alpha_{s,zz}^{em}}{\sqrt{\varepsilon_0 \mu_0}} e^{i\psi}\right)^* \frac{-3}{2\pi d^4} |\text{E}_0|^2 \alpha_t^{ee}\right] . \tag{13}$$

Therefore, the force difference between any two phase-shift parameters $\psi_1$ and $\psi_2$ reads

$$F_z^{\text{ARPB}}\big|_{\psi_2} - F_z^{\text{ARPB}}\big|_{\psi_1} = -\frac{3F_Y |\text{E}_0|^2}{4\pi\sqrt{\varepsilon_0 \mu_0} d^4}\left[\text{Re}\left(e^{-i\psi_2}\alpha_t^{ee}\alpha_{s,zz}^{em\,*}\right) - \text{Re}\left(e^{-i\psi_1}\alpha_t^{ee}\alpha_{s,zz}^{em\,*}\right)\right], \tag{14}$$

Now, we prove that the maximum force difference for any choice of the two phase parameters $\psi_1$ and $\psi_2$ occurs when $F_Y^2 = |V / (\eta_0 I)|^2 = 1$ , i.e., when the illumination powers of the APB and RPB (with the same beam waist) are equal. To this end, we assume the amplitude coefficient of the RPB, defined in the manuscript, to be $I$ and the amplitude coefficient of APB, defined in the manuscript, to be $V = F_Y I / \eta_0$ . According to Eqs. (14) and (17) of the manuscript, the total ARPB power reads

$$P_{\text{tot}} = F_Y^2 \frac{|I|^2}{2} \eta_0 \left(1 - \frac{1}{2(\pi w_0 / \lambda)^2}\right) + \frac{|I|^2}{2} \eta_0 \left(1 - \frac{1}{2(\pi w_0 / \lambda)^2}\right). \tag{15}$$

Under the above assumption the force difference between two measurements for ARPB illuminations with two different phase parameters $\psi_1$ and $\psi_2$ in Eq. (14) reads

$$F_z^{\text{ARPB}}(\psi_2) - F_z^{\text{ARPB}}(\psi_1) = -\frac{12F_Y \eta_0 |I|^2}{\pi^2 w_0^4 \omega^2 \sqrt{\varepsilon_0 \mu_0} d^4 \varepsilon_0^2}\left[\text{Re}\left(e^{-i\psi_2}\alpha_t^{ee}\alpha_{s,zz}^{em\,*}\right) - \text{Re}\left(e^{-i\psi_1}\alpha_t^{ee}\alpha_{s,zz}^{em\,*}\right)\right]. \tag{16}$$

where

$$|I|^2 = \frac{1}{(1 + F_Y^2)} \frac{2P_{\text{tot}}}{\left(1 - \frac{1}{2(\pi w_0 / \lambda)^2}\right)}, \tag{17}$$

from Eq.(15). To find the value of $F_Y$ for which the differential force in Eq. (16) peaks, assuming a constant $P_{\text{tot}}$ , we replace $|I|^2$ from Eq. (17) into Eq. (16) and set the derivative of Eq. (16) with respect to $F_Y$ equal to zero and get

$$\frac{1}{F_Y^2 + 1} - \frac{2F_Y^2}{\left(F_Y^2 + 1\right)^2} = 0. \tag{18}$$

For this equation to hold, we require to have $F_Y = 1$.

Let us now investigate the maximum force difference for any arbitrary $F_Y$, i.e., the swing as defined in Eq. (19) of the manuscript. To this end, we expand Eq. (14) as

$$\Delta F_z = -\frac{3F_Y |\text{E}_0|^2}{4\pi\sqrt{\varepsilon_0 \mu_0} d^4}\left[\left(\cos\psi_2 - \cos\psi_1\right)\text{Re}\left(\alpha_t^{ee}\alpha_{s,zz}^{em\,*}\right) + \left(\sin\psi_2 - \sin\psi_1\right)\text{Im}\left(\alpha_t^{ee}\alpha_{s,zz}^{em\,*}\right)\right]. \tag{19}$$

As it is observed from this equation, since $\alpha_t^{ee} \alpha_{s,zz}^{em\,*}$ has a complex value for a general lossy particle, then, it is not straightforward to conclude that the swing happens when $|\psi_2 - \psi_1| \approx 180$. However, as we have shown in the manuscript (see e.g. Fig. 5(b) of the manuscript) our numerical analyses demonstrate such a conclusion and we analytically prove it here for special case of particles with negligible losses.

Assuming tip and sample to have negligible losses (under this assumption $\alpha_s^{em}$ is purely imaginary) Eq. (19) reduces to

$$\Delta F_z \approx -\frac{3F_Y |\mathrm{E}_0|^2}{4\pi \sqrt{\varepsilon_0 \mu_0}\, d^4} \mathrm{Re}\left(\alpha_t^{ee}\right)\mathrm{Im}\left(\alpha_{s,zz}^{em}\right)^* \left(\sin\psi_2 - \sin\psi_1\right). \tag{20}$$

Since $|\sin\psi| < 1$, then the maximum value of Eq. (20) occurs when $|\psi_2 - \psi_1| = 180$ which results in $|\sin\psi_2 - \sin\psi_1| = 2$. Therefore, the maximum swing force, i.e., the difference between the maximum and minimum of exerted force on the tip-apex $F_z$, when varying $\psi_1$ and $\psi_2$, , defined in Eq. (19) of the manuscript, is obtained from Eq. (20) when $F_Y = 1$ and $|\sin\psi_2 - \sin\psi_1| = 2$ , and the final result is given as Eq. (22) of the manuscript.

## REFERENCES

(1)    Kamandi, M.; Albooyeh, M.; Guclu, C.; Veysi, M.; Zeng, J.; Wickramasinghe, K.; Capolino, F. Enantiospecific Detection of Chiral Nanosamples Using Photoinduced Force. *Phys. Rev. Appl.* **2017**, *8* (6), 064010.
(2)    Guclu, C.; Tamma, V. A.; Wickramasinghe, H. K.; Capolino, F. Photo-Induced Magnetic Force Between Nanostructures. *Phys. Rev. B* **2015**, *92* (23).